\documentclass[
showpacs,preprintnumbers,amsmath,amssymb]{revtex4}
\usepackage{amsmath}

\begin{document}

\tolerance=5000

\def\pp{{\, \mid \hskip -1.5mm =}}
\def\cL{{\cal L}}
\def\be{\begin{equation}}
\def\ee{\end{equation}}
\def\bea{\begin{eqnarray}}
\def\eea{\end{eqnarray}}
\def\beq{\begin{eqnarray}}
\def\eeq{\end{eqnarray}}
\def\tr{{\rm tr}\, }
\def\nn{\nonumber \\}
\def\e{{\rm e}}

\title{Modified non-local-F(R) gravity as the key for the inflation and
dark energy}
\author{Shin'ichi Nojiri}
\email{nojiri@phys.nagoya-u.ac.jp}
\affiliation{Department of Physics, Nagoya University, Nagoya 464-8602. Japan}
\author{Sergei D. Odintsov\footnote{also at Lab. Fundam. Study, Tomsk State
Pedagogical University, Tomsk}}
\email{odintsov@ieec.uab.es}
\affiliation{Instituci\`{o} Catalana de Recerca i Estudis Avan\c{c}ats (ICREA)
and Institut de Ciencies de l'Espai (IEEC-CSIC),
Campus UAB, Facultat de Ciencies, Torre C5-Par-2a pl, E-08193 Bellaterra
(Barcelona), Spain 
and YITP, Kyoto University, Kyoto, Japan}
\preprint{YITP-07-49}

\begin{abstract}

We consider FRW cosmology in non-local modified gravity. Its local
scalar-tensor 
formulation  is developed. It is explicitly demonstrated
that such theory  may lead to the
unification of early-time inflation with late-time cosmic acceleration. 
The quintessence or phantom era may  emerge for specific form of the
action.
The coupled non-local-$F(R)$ gravity is also investigated. It is shown
that
such theory being consistent with Solar System tests may lead to the known 
universe history sequence: inflation, radiation/matter dominance and
dark epoch.

\end{abstract}


\maketitle

\section{Introduction}

Modified gravity \cite{review} suggests serious 
alternative for dark
energy origin. Indeed, it may be naturally expected that
gravitational action 
contains some extra terms which became relevant recently with the
significant decrease of the universe curvature. The number of metric
formulation modified $F(R)$ 
gravities has been proposed \cite{review,CDTT,NO,FR,FR1} with the
purpose to explain the origin of cosmic acceleration. Special attention is
paid to $F(R)$ models \cite{cap,HS,Uf,AB} with the effective cosmological
constant phase because such theories may easily reproduce the well-known
$\Lambda$CDM cosmology. Such models subclass which does not violate Solar
System tests represents the real alternative for standard General
Relativity.

From another point, it is expected that due to string/M-theory corrections 
the early-time universe may be also governed by modified gravity which
initiates the inflationary epoch. Some (phenomenological) modified
gravities which may unify the inflation and dark energy have been
proposed \cite{NO,Uf}. The search for viable gravitational dark
energy/inflation
continues. Recently, it was suggested to consider non-local
gravity\cite{deser} which may be produced by quantum effects as
the source of cosmic acceleration. 

In the present letter we propose the class of non-local modified
gravities with the effective cosmological constant epoch. The
local scalar-tensor formulation of such theory is
developed. Several explicit examples are considered where it is shown that 
such non-local gravity may lead to the unification of the early-time
inflation with late-time acceleration being consistent with the known 
universe expansion history and local tests. For other examples, non-local
 modified gravity may lead to quintessence/phantom cosmology.
Non-local gravity coupled with $F(R)$ theory is considered. It is
demonstrated that it is even easier to achieve the unification of the
early-time inflation with late-time acceleration (with radiation/matter
dominance phases between them) in such coupled theory which seems to be
consistent with local tests. In fact, it is shown that it is natural to
unify the early-time inflation with late-time
acceleration in modified gravity with two different types of terms:
first being relevant at the early universe and second being relevant at
late universe.

\section{Unifying inflation with dark energy in the non-local modified
gravity}

In this section we present local (scalars-tensor) formulation of
non-local gravity. The explicit example of such theory which naturally
leads to the unification of inflation with cosmic acceleration is 
worked out.

The starting action of the non-local gravity is given by
\be
\label{nl1}
S=\int d^4 x \sqrt{-g}\left\{
\frac{1}{2\kappa^2}R\left(1 + f(\Box^{-1}R)\right) + {\cal L}_{\rm matter} 
\right\}\ .
\ee
Here $f$ is some function and $\Box$ is the d'Almbertian for scalar field.
Note that our approach is purely phenomenological. Generally speaking,
such non-local effective action may be induced by quantum effects (for
instance, via RG improving \cite{eli}). 
The above action can be rewritten by introducing two scalar fields $\phi$
 and $\xi$ in the following form:
\bea
\label{nl2}
S&=&\int d^4 x \sqrt{-g}\left[
\frac{1}{2\kappa^2}\left\{R\left(1 + f(\phi)\right) 
+ \xi\left(\Box\phi - R\right) \right\}
+ {\cal L}_{\rm matter} 
\right] \nn
&=&\int d^4 x \sqrt{-g}\left[
\frac{1}{2\kappa^2}\left\{R\left(1 + f(\phi)\right) 
 - \partial_\mu \xi \partial^\mu \phi - \xi R \right\}
+ {\cal L}_{\rm matter} 
\right]
\ .
\eea
By the variation over $\xi$, we obtain
$\Box\phi=R$ or $\phi=\Box^{-1}R$.
Substituting the above equation into (\ref{nl2}), one reobtains
(\ref{nl1}). 

Varying (\ref{nl2}) with respect to the metric tensor $g_{\mu\nu}$ gives
\bea
\label{nl4}
0 &=& \frac{1}{2}g_{\mu\nu} \left\{R\left(1 + f(\phi) - \xi\right) 
 - \partial_\rho \xi \partial^\rho \phi \right\}
 - R_{\mu\nu}\left(1 + f(\phi) - \xi\right) \nn
&& + \frac{1}{2}\left(\partial_\mu \xi \partial_\nu \phi + \partial_\mu \phi \partial_\nu \xi \right) 
 -\left(g_{\mu\nu}\Box - \nabla_\mu \nabla_\nu\right)\left( f(\phi) - \xi\right) 
+ \kappa^2T_{\mu\nu}\ .
\eea
On the other hand, the variation with respect to $\phi$ gives
\be
\label{nl5}
0=\Box\xi+ f'(\phi) R\ .
\ee
Now we assume the FRW metric 
\be
\label{nl6}
ds^2 = - dt^2 + a(t)^2 \sum_{i=1,2,3}\left(dx^i\right)^2\ ,
\ee
and the scalar fields $\phi$ and $\xi$ only depend on time. 
Then Eq.(\ref{nl4}) has the following form:
\bea
\label{nl7a}
0 &=& - 3 H^2\left(1 + f(\phi) - \xi\right) + \frac{1}{2}\dot\xi \dot\phi - 3H\left(f'(\phi)\dot\phi - \dot\xi\right) 
+ \kappa^2 \rho\ ,\\
\label{nl7b}
0 &=& \left(2\dot H + 3H^2\right) \left(1 + f(\phi) - \xi\right) + \frac{1}{2}\dot\xi \dot\phi 
+ \left(\frac{d^2}{dt^2} + 2H \frac{d}{dt} \right) \left( f(\phi) - \xi \right) + \kappa^2 p\ .
\eea
On the other hand, scalar equations are:
\bea
\label{nl8a}
0 &=& \ddot \phi + 3H \dot \phi + 6 \dot H + 12 H^2 \ , \\
\label{nl8b}
0 &=& \ddot \xi + 3H \dot \xi - \left( 6 \dot H + 12 H^2\right)f'(\phi) \ .
\eea

The remark is in order.
We may consider more general action:
\be
\label{nl9}
S=\int d^4 x \sqrt{-g} \left[ F\left(R, \Box R, \Box^2 R , \cdots, \Box^m R, \Box^{-1} R , \Box^{-2} R , 
\cdots , \Box^{-n} R \right) + {\cal L}_{\rm matter} \right]\ .
\ee
Here $m$ and $n$ are positive integers. Again, for the action (\ref{nl9}),
by introducing $2n$-scalars, 
one can rewrite the action (\ref{nl9}) in a local form:
\be
\label{nl10}
S=\int d^4 x \sqrt{-g} \left[ F\left(R, \Box R, \Box^2 R , \cdots, \Box^m R, \phi_1, \phi_2, 
\cdots ,\phi_n \right) 
+ \sum_{k=1}^n \xi_k \left(\Box^k \phi_k - R \right) + {\cal L}_{\rm matter} \right]\ .
\ee
The generalization for non-integer $m$ is also possible.

We now assume deSitter solution $H=H_0$, then Eq.(\ref{nl8a}) can be solved as
\be
\label{NLdS1}
\phi= - 4H_0 - \phi_0 \e^{-3H_0 t} + \phi_1\ ,
\ee
with constants of integration, $\phi_0$ and $\phi_1$. 
For simplicity, we only consider the case that $\phi_0=\phi_1=0$. 
We also assume $f(\phi)$ is given by
\be
\label{NLdS2}
f(\phi)=f_0 \e^{b\phi}= f_0 \e^{-4bH_0 \phi}\ .
\ee
Then Eq.(\ref{nl8b}) can be solved as follows,
\be
\label{NLdS3}
\xi= - \frac{3f_0}{3 - 4b} \e^{-4bH_0 t} + \frac{\xi_0}{3H_0}\e^{-3H_0 t} - \xi_1\ .
\ee
Here $\xi_0$ and $\xi_1$ are constants. 
For the deSitter space $a$ behaves as $a=a_0\e^{H_0 t}$. Then for the matter with constant equation of state $w$, 
we find
\be
\label{NN1}
\rho = \rho_0 \e^{-3(w+1)H_0 t}\ .
\ee
Then by substiruting (\ref{NLdS1}), (\ref{NLdS3}), and (\ref{NN1}) into (\ref{nl7a}), we obtain
\be
\label{NLdS4}
0 = - 3H_0^2 \left(1 + \xi_1\right) + 6H_0^2 f_0 \left( 2b - 1 \right)\e^{-4H_0b t}
 + \kappa^2  \rho_0 \e^{-3(w+1)H_0 t} \ .
\ee
When $\rho_0=0$, if we choose
\be
\label{NLdS5}
b=\frac{1}{2}\ ,\quad 
\xi_1 = - 1 \ ,
\ee
deSitter space can be a solution. 
Even if $\rho\neq 0$, if we choose
\be
\label{NN2}
b=\frac{3}{4}(1+w) \ ,\quad f_0 = \frac{\kappa^2 \rho_0}{3H_0^2 \left(1 + 3w \right)}\ ,\quad \xi_1 = -1\ ,
\ee
there is a deSitter solution. 

In the presence of matter with $w\neq 0$, we may have a deSitter solution $H=H_0$ even if $f(\phi)$ 
given by 
\be
\label{nLL1}
f(\phi)=f_0\e^{\phi/2} + f_1\e^{3(w+1)\phi/4}\ .
\ee
Then the following solution exists:
\be
\label{nLL2}
\phi= - 4H_0 t\ ,\quad \xi=1 + 3f_0\e^{-2H_0t}+ \frac{f_1}{w}\e^{-3(w+1)H_0 t}\ , \quad 
\rho= - \frac{3(3w+1)H_0^2 f_1}{\kappa^2}\e^{-3(1+w)H_0t}\ .
\ee

Note that $H_0$ in (\ref{NLdS1}) can be arbitrary and can be determined by an initial
condition. Since $H_0$ can be small or large, the theory with function {NLdS2} with $b=1/2$ 
could describe the early-time inflation  or current cosmic acceleration. 
Motivated by this, we may propose the following model:
\be
\label{nl13}
f(\phi)=\left\{ \begin{array}{lc}
f_0\e^{\phi/2}\quad &0>\phi>\phi_1 \\ 
f_0\e^{\phi_1/2} \quad &\phi_1 > \phi > \phi_2 \\
f_0\e^{\left(\phi - \phi_2 + \phi_1\right) /2} \quad & \phi < \phi_2 
\end{array} \right. \ .
\ee
Here $\phi_1$ and $\phi_2$ are constants. 
We also assume that matter could be neglected when $0>\phi>\phi_1$ or $\phi < \phi_2$. 
Since the above function $f(\phi)$ is not smooth around $\phi=\phi_1$ and $\phi_2$, 
one may replace the above $f(\phi)$ with a more smooth function. 
When $0>\phi>\phi_1$ or $\phi < \phi_2$, the universe is described by the deSitter solution  
although corresponding $H_0$ might be different. 
When $\phi_1 > \phi > \phi_2$, since $f(\phi)$ is a constant, the universe is described 
by the Einstein gravity, where effective gravitational constant $\kappa_{\rm eff}$ 
is given by
\be
\label{nl14}
\frac{1}{\kappa_{\rm eff}^2}=\frac{1}{\kappa^2}\left( 1 + f_0\e^{\phi_1/2} \right)\ .
\ee
(Note that in non-local gravity when auxiliary scalar is not constant,
the Newton coupling constant is defined ambigiously.)
Then due to the matter contribution there could occur matter dominated
phase. 
In this phase, the Hubble rate $H$ behaves as $H=\frac{2}{3\left(t_0 + t\right)}$
with a constant $t_0$ and the scalar curvature is given by $R=\frac{4}{3\left(t_0 + t\right)^2}$.
Now we assume that the universe started at $t=0$ with a rather big but constant curvature $R=R_I= 12 H_I^2$ 
with a constant $H_I$, that is, the universe is in deSitter phase.
Then in the model (\ref{nl13}), by  following (\ref{NLdS1}), $\phi$ behaves as $\phi=- 4H_I t$. 
Subsequently, at $t=t_1\equiv - \phi_1/4H_I$, we have $\phi=\phi_1$ 
and the universe enters into the matter dominated phase. If the curvature
is continuous at $t=t_1$, $t_0$ can be found by solving 
\be
\label{nl17}
R=\frac{4}{3\left(t_0 + t_1\right)^2} = 12 H_I^2 \ .
\ee
If $\phi$ and $\dot\phi$ are also continuous, when $\phi_1 > \phi > \phi_2$,
$\phi$ is given by solving (\ref{nl8a}) as
\be
\label{nl18}
\phi = - \frac{4}{3}\ln \left(\frac{t}{t_1}\right) - \tilde\phi \left(t - t_1\right) + \phi_1\ ,\quad
\tilde\phi \equiv - 4H_I \left(t_0 + t_1\right)^2 + \frac{4}{3}\left(t_0 + t_1\right)\ .
\ee
When $\phi=\phi_2$, the deSitter phase, which corresponds to the accelerating 
expansion of the present universe, could have started. 
The solution corresponds to deSitter space 
(with some shifts of parameters) and $H_0=H_L$ could be given by solving
\be
\label{nl19}
12 H_L^2 = \frac{4}{3\left(t_0 + t_2\right)^2}\ .
\ee
if the curvature is continuous at $\phi=\phi_2$. 
In (\ref{nl19}), $t_2$ is defined by $\phi(t_2)=\phi_2$. 
Thus, we got  the cosmological FRW model with inflation, radiation/matter
dominated phase, and current accelerating expansion.

\section{Phantom cosmology in the non-local gravity}

Let us demonstrate that non-local gravity may also lead to the effective 
phantom cosmology without need to introduce the non-physical
scalar with negative kinetic energy. The quintessence cosmology may also emerge.

We now investigate if there could be power law solution corresponding to 
quintessence or phantom cosmology, where 
\be
\label{NN3}
H=\frac{h_0}{t} \quad \left(a\propto t^{h_0}\right)\ .
\ee
Then a solution of Eq.(\ref{nl8a}) is given by
\be
\label{NN4}
\phi=\phi_0 \ln \frac{t}{t_0}\ ,\quad 
\phi_0= \frac{-6h_0 + 12 h_0^2}{1 - 3h_0}\ .
\ee
Here $t_0$ is a constant. 
We now assume, as in (\ref{NLdS2}), $f(\phi)$ is given by
\be
\label{NN5}
f(\phi)=f_0 \e^{b\phi}= f_0 \left(\frac{t}{t_0}\right)^{b h_0}\ .
\ee
Then by solving Eq.(\ref{nl8b}) as
\be
\label{NN6}
\xi = \frac{\left( - 6 h_0 + 12 h_0^2 \right)f_0}{\left(b\phi_0 + 3h_0 -1\right)\phi_0}\left(\frac{t}{t_0}\right)^{b h_0} 
+ \frac{t_0 \tilde\xi_0}{-3h_0 + 1}\left(\frac{t}{t_0}\right)^{ - 3 h_0 + 1} + \tilde\xi_1\ .
\ee
Here $\tilde\xi_0$ and $\tilde\xi_1$ are constants of integration. 
We also find 
\be
\label{NN7}
\rho= \tilde\rho_0 \left(\frac{t}{t_0}\right)^{-3(1+w)h_0}\ .
\ee
When $\rho_0=0$, Eq. (\ref{nl7a}) can be satisfied if
\be
\label{NN8}
\tilde\xi_1=1\ ,\quad 
0= - 6h_0^2\left( - 6 h_0 + 12 h_0^2 \right)^2 b^2 
+ \left(-42 h_0^2 + 9h_0 + 1\right)\left( - 6 h_0 + 12 h_0^2 \right)\left(1-3h_0\right) b 
+ 6h_0^2\left(1-3h_0\right)^3\ .
\ee
$\tilde\xi_0$ and $f_0$ could be arbitrary. Then if we give $b$ satisfying the second equation in (\ref{NN8}), 
there could be a power law solution. 
Even if $\rho_0\neq 0$, there could be a solution if
\be
\label{NN9}
\tilde\xi_1=1\ ,\quad 
b=\frac{\left\{2 - 3\left(1+w\right)h_0\right\}\left(1 - 3h_0\right)}{-6h_0^2 + 12 h_0^2}\ ,\quad
f_0 = - \frac{\kappa^2 \rho_0}{-3h_0^2 + 3 h_0 + 9 h_0 w 
+ \frac{\left(-6 h_0^2 - \frac{9}{2} h_0 - 3 \left(3h_0 + \frac{1}{2}\right)h_0w\right) \left(1 - 3h_0\right)}{1 - 3 h_0 w}}\ .
\ee
In other words, if we start with the theory where $b$ and $f_0$ are given by (\ref{NN9}), 
we obtain the cosmological solution given by (\ref{NN3}) with (\ref{NN4}) and (\ref{NN6}). 
In the above formulation $h_0$ is an arbitrary. Since the effective 
equation of state (EoS) parameter $w_{\rm eff}$ is given by 
\be
\label{nl26}
w_{\rm eff}\equiv -1 - \frac{2\dot H}{3H^2} = - 1 + \frac{2}{3h_0}\ ,
\ee
any $w_{\rm eff}$ corresponding to quintessence or phantom can be realized
in this non-local gravity. Especially if $h_0$ is negative, the effective phantom 
cosmology occurs. When $h_0<0$, we may shift $t$ as $t - t_s$ and assume $t<t_s$ in the present unverse. 
Then we have 
\be
\label{nl27}
H=-\frac{h_0}{t_s - t}\ ,
\ee
and $t=t_s$ corresponds to the Big Rip singularity. 

\section{Unification of the inflation with cosmic acceleration in
the non-local-F(R) gravity}

Let us discuss the accelerating early-time and late-time cosmology in the
non-local gravity where $F(R)$-term \cite{review} is added. The starting action is: 
\be
\label{nl30}
S=\int d^4 x \sqrt{-g}\left\{
\frac{1}{2\kappa^2}R\left(1 + f(\Box^{-1}R)\right) + F(R) + {\cal L}_{\rm matter} 
\right\}\ .
\ee
Here $F(R)$ is some function of $R$. FRW equations look like
\bea
\label{nl31}
0 &=& - 3 H^2\left(1 + f(\phi) - \xi\right) + \frac{1}{2}\dot\xi \dot\phi - 3H\left(f'(\phi)\dot\phi - \dot\xi\right) \nn
&& - F(R) + 6\left(H^2 + \dot H\right) F'(R) - 36\left(4H^2 \dot H + H \ddot H\right)F''(R) 
+ \kappa^2 \rho\ ,\\
\label{nl32}
0 &=& \left(2\dot H + 3H^2\right) \left(1 + f(\phi) - \xi\right) + \frac{1}{2}\dot\xi \dot\phi 
+ \left(\frac{d^2}{dt^2} + 2H \frac{d}{dt} \right) \left( f(\phi) - \xi \right) \nn 
&& F(R) - 2\left(\dot H + 3H^2\right)F'(R) + \kappa^2 p\ .
\eea
Here $R=12H^2 + 6\dot H$.

We may propose several scenarios. One is that the inflation at the early
universe is generated mainly by 
$F(R)$ part but the current acceleration is defined  
mainly by $f\left(\Box^{-1}R\right)$ part. One may consider the inverse,
that is, the inflation is generated by 
$f\left(\Box^{-1}R\right)$ part but the late-time acceleration 
by $F(R)$. 

For instance, for the first scenario one can take:
$F(R)= \beta R^2$.
Here $\beta$ is a constant. We choose $f(\Box^{-1}R)$ part as in
(\ref{NLdS2}) with $b=1/2$ but $f_0$ is taken to be very small 
and $\phi$ starts with $\phi=0$. Hence, at the early universe
$f\left(\Box^{-1}R\right)$ is very small and 
could be neglected. Then due to the $F(R)$-term (\ref{nl32}),
there occurs (slightly modified) $R^2$-inflation. After the end of the inflation, 
there occurs the radiation/matter dominance era. In this phase, $\phi$
behaves as in (\ref{nl18}):
$\phi = - \frac{4}{3}\ln \left(\frac{t}{\hat t_0}\right) - \hat\phi_1
\left(t - \hat t_0\right) + \hat\phi_2$. 
However, the constants $\hat t_0$, $\hat\phi_1$, and $\hat\phi_2$ should
be determined by the proper 
initial conditions, which may differ from that in (\ref{nl18}). We now
assume $\hat\phi_1$ is very small but negative. 
From the expression of (\ref{NLdS2}) it follows $f(\phi)$ becomes large as
time goes by and finally this term 
dominates. As a result, deSitter expansion  occurs at
the present universe. 

For the second scenario, the early-time inflation 
is generated by 
$f\left(\Box^{-1}R\right)$ part but the cosmic acceleration is generated  
by $F(R)$. 
As an $F(R)$-term, one can take the model \cite{HS}:
\be
\label{HS1}
F_{HS}(R)=-\frac{m^2 c_1 \left(R/m^2\right)^n}{c_2 \left(R/m^2\right)^n + 1}\ ,
\ee
which has the following properties
\bea
\label{HS2}
\lim_{R\to\infty} F_{HS} (R) &=& \mbox{const}\ ,\nn
\lim_{R\to 0} F_{HS}(R) &=& 0\ ,
\eea
The second condition means that there is a flat spacetime solution
(vanishing cosmological constant).
The estimation of ref.\cite{HS} suggests that $R/m^2$ is not so small but rather large even
at the present universe and $R/m^2\sim 41$.
Hence,
$F_{HS}(R)\sim - \frac{m^2 c_1}{c_2} + \frac{m^2 c_1}{c_2^2}
\left(\frac{R}{m^2}\right)^{-n}$,
which gives an ``effective'' cosmological constant $-m^2 c_1/c_2$ and
generates the late-time accelerating expansion.
One can show that
\be
\label{HSbb1}
H^2 \sim \frac{m^2 c_1 \kappa^2 }{c_2} \sim \left(70 \rm{km/s\cdot pc}\right)^2 \sim \left(10^{-33}{\rm eV}\right)^2\ .
\ee
At the intermediate epoch, where the matter density $\rho$ is larger
than the effective cosmological constant,
$\rho > \frac{m^2 c_1}{c_2}$,
there appears the matter dominated phase and the universe
expands with deceleration. Hence, above model describes the effective
$\Lambda$CDM cosmology.

As a $f\left(\Box^{-1}R\right)$ part, we consider theory (\ref{NLdS2}) with $b=1/2$,
again. 
It is assumed $f_0$ is large and $f\left(\Box^{-1}R\right)$ term could be
dominant at the early universe. 
Hence, following (\ref{NLdS1}), $\phi$ becomes negative and large as time
goes by and therefore 
$f(\phi)$ becomes small and could be neglected at late universe. Then
there appears naturally the radiation/matter dominated phase. 
After that due to $F_{HS}(R)$-term (\ref{HS1}),
the late-time acceleration occurs.

Although the model \cite{HS} is quite succesful, the early
time inflation is not included there. In \cite{Uf}, it was suggested 
the modified gravity
model to treat the inflation and the late-time accelerating expansion
 in a unified way.
In order to generate the inflation, one may require
$\lim_{R\to\infty} f (R) = - \Lambda_i$.
Here $\Lambda_i$ is an effective cosmological constant at the early
universe and therefore we assume $\Lambda_i \gg \left(10^{-33}{\rm eV}\right)^2$.
One may assume $\Lambda_i\sim 10^{20\sim38}$.
In order that the current accelerating expansion  could be
generated, 
let us consider that $f(R)$ is a small constant at present universe, that
is,
$f(R_0)= - 2R_0$, $f'(R_0)\sim 0$.
Here $R_0$ is current curvature $R_0\sim \left(10^{-33}{\rm eV}\right)^2$.
The next condition corresponding to the second one in (\ref{HS2}) is:
$\lim_{R\to 0} f(R) = 0$.
In the above class of models, the early universe starts from the
inflation
driven by the effective cosmological constant. 
As curvature becomes smaller, the
effective cosmological constant also becomes smaller. After that the
radiation/matter dominates. When the density of the radiation and the matter 
becomes small and the curvature goes to the value $R_0$,
there appears the small effective cosmological constant.
Hence, the current cosmic expansion starts.

In \cite{Uf}, two examples have been proposed.
The first model is given by
\be
\label{Uf5}
F(R) = - \frac{\left(R-R_0\right)^{2n+1} + R_0^{2n+1}}{f_0
+ f_1 \left\{\left(R-R_0\right)^{2n+1} + R_0^{2n+1}\right\}}
=- \frac{1}{f_1} + \frac{f_0/f_1}{f_0
+ f_1 \left\{\left(R-R_0\right)^{2n+1} + R_0^{2n+1}\right\}}\ .
\ee
Here $n$ is a positive integer, $n=1,2,3,\cdots$ and
\be
\label{Uf6}
\frac{R_0^{2n+1}}{f_0 +f_1 R_0^{2n+1}}=2R_0\ ,\quad
\frac{1}{f_1} = \Lambda_i\ ,
\ee
that is
\be
\label{Uf7}
f_0=\frac{R_0^{2n}}{2} - \frac{R_0^{2n+1}}{\Lambda_i}
\sim \frac{R_0^{2n}}{2} \ ,\quad
f_1=\frac{1}{\Lambda_i}\ .
\ee
The second model is 
\be
\label{Uf16}
F(R)=-f_0 \int_0^R dR \e^{-\frac{\alpha R_1^{2n}}{\left(R - R_1\right)^{2n}} - \frac{R}{\beta\Lambda_i}}\ .
\ee
Here $\alpha$, $\beta$, $f_0$, and $R_1$ are constants.
Then by construction, as long as $0<f_0<1$, $F'(R)>-1$, which shows that
there is no anti-gravity regime. 
Since
\bea
\label{Uf17}
f(R_1) &\sim& -f_0 \int_0^{R_1} dR \e^{-\frac{\alpha R_1^{2n}}{\left(R - R_1\right)^{2n}}}
= -f_0 A_n(\alpha) R_1\ , \nn
A_n(\alpha) &\equiv& \int_0^1 dx \e^{-\frac{\alpha}{x^{2n}}}\ ,
\eea
and $- f(R_1)$ could be identified with the effective cosmological constant $2R_0$, we find
\be
\label{Uf18}
f_0 A_n(\alpha) R_1 = R_0\ .
\ee
Note that $A_n(0)=1$, $A_n(+\infty)=0$, and $A'(x)<0$.
On the other hand, since
\be
\label{Uf19}
f(+\infty) \sim \int_0^{\infty}dR \e^{-{R}{\beta \Lambda_i}} = - f_0 \beta \Lambda_i\ ,
\ee
and $-f(+\infty)$ could be identified with the effective cosmological constant 
at the inflationary epoch, $\Lambda_i$, one gets $f_0\beta=1$.

As a $f\left(\Box^{-1}R\right)$ part, we consider (\ref{NLdS2}) with $b=1/2$, again and
assume $f_0$ is large. 
Thus, $f\left(\Box^{-1}R\right)$ term could be dominant at the early
universe. 
Then the inflation is generated by the combination of $F(R)$ with
non-local term.
Following (\ref{NLdS1}), $\phi$ becomes negative and large as time goes by
and therefore 
$f(\phi)$ becomes small and could be neglected at late
universe. After that, there
occurs the radiation/matter dominated phase. 
After that due to $F_{HS}(R)$-term (\ref{HS1}),
the accelerating expansion starts.

Some nice features of above scenario are related with Newton law which
should be respected at current universe. 
For the $F(R)$-models (\ref{HS1}), (\ref{Uf5}), and (\ref{Uf16}),
the Newton law corrections have been found in 
\cite{Uf}. It was shown that currently the corrections are very small. 
In the above scenario, where the inflation occurs due to
$f\left(\Box^{-1}R\right)$ part, its contribution to FRW dynamics 
becomes very small at late universe. Hence, the contribution from
$f\left(\Box^{-1}R\right)$ part to 
the Newton law correction is negligible. 
The same is true for so-called matter instability\cite{DK,NO,Faraoni}, 
which happens in some models of $F(R)$-gravity. It has been shown that
such instability is absent 
in the above $F(R)$-models (\ref{HS1}), (\ref{Uf5}), and
(\ref{Uf16}) \cite{Uf}. Then since the contribution 
from $f\left(\Box^{-1}R\right)$ is very small at the late time universe,
there is no such
instability in above non-local $F(R)$ gravity at late universe.
Thus, we demonstrated that non-local gravity coupled with $F(R)$-term 
may naturally predict the known universe expansion sequence: inflation,
radiation/matter dominance and dark energy.

\section{Discussion}

In summary, we demonstrated that non-local gravity may be a key for the
origin of the inflation and dark energy being consistent with simplest
local tests and the known universe expansion history. The known universe 
epochs sequence is easier to realize when modified gravity is given as
some combination of terms where one term generates the inflation while
other, qualitatively different term pushes the late universe to
accelerate. In addition, such combined modified gravity easily passes 
the simplest local tests. The corresponding example of non-local-$F(R)$
gravity is discussed in detail. 

More complicated versions of non-local
gravity may be considered in similar scalar-tensor formulation with
auxiliary scalars. It is important that even if more precise observational
data  define the EoS parameter $w$ to be slightly different from $-1$,
there exists the possibility to realize such scenario in non-local gravity
as the effective quintessence or phantom cosmology. Moreover, the
cosmological perturbations (for a review, see \cite{sasaki}) should be
investigated there. This will be discussed elsewhere. 

\section*{Acknowledgements}

This work started from the meeting in JSPS-DST collaboration. 
We thank  M. Sasaki, S. Tsujikawa, S. Panda, T. Shiromizu, 
and S. Mizuno for helpful discussions. Special thanks are given to M. Sami
for the participation at the early stage of this work. 
The research by S.N. has been supported in part by the
Ministry of Education, Science, Sports and Culture of Japan under 
grant no.18549001 and 21st Century COE Program of Nagoya University
provided by Japan Society for the Promotion of Science (15COEG01).
The research by S.D.O. has been supported in part by the projects
FIS2006-02842 (MEC, Spain), by the project 2005SGR00790
(AGAUR,Catalunya) and especially, by YITP, Kyoto University.

\appendix

\section{Stability in deSitter background}

We now check the (in)stability in the deSitter solution in (\ref{NLdS1}-\ref{NLdS5}). 
For simplicity, we consider the case without matter.  
By defining
\be
\label{A1}
X\equiv - \frac{\dot \phi}{4H}\ ,\quad Y \equiv \frac{1-\xi}{3f(\phi)}\ ,\quad 
W=\frac{\dot \xi}{6H f(\phi)}\ ,\quad \frac{d}{dN}\equiv a\frac{d}{da} = \frac{1}{H}\frac{d}{dt}\ ,
\ee
Eqs.(\ref{nl7a}), (\ref{nl8a}), and (\ref{nl8b}) could be rewritten as
\bea
\label{A2}
0 &=& - (1+3Y) - 4XW + 2X + 6W\ ,\\
\label{A3}
\frac{dX}{dN} &=& - \left(\frac{1}{H}\frac{dH}{dN} + 3\right)X 
+ \frac{3}{2} \left(\frac{1}{H}\frac{dH}{dN} + 2\right)\ , \\
\label{A4}
\frac{dW}{dN} &=& - \left(\frac{1}{H}\frac{dH}{dN} + 3\right)W 
+ \frac{1}{2}\left(\frac{1}{H}\frac{dH}{dN} + 2\right) + 2WX\ .
\eea
For the deSitter solution in (\ref{NLdS1}-\ref{NLdS5}), we have
\be
\label{A4b}
X=Y=W=1\ .
\ee
By multiplying $d/dN$ with (\ref{A2}) and using (\ref{A3}) and (\ref{A4}) and 
deleting $Y$ by (\ref{A2}), we obtain
\be
\label{A5}
0= \left( 2 - X - 6W + 4WX\right)\frac{1}{H}\frac{dH}{dN} 
+ 6 - 12 W - 4 X - 2 X^2 + 12 WX\ .
\ee
We now consider the perturbation from the deSitter solution (\ref{A4b}):
\be
\label{A6}
X=1 + \delta X\ ,\quad W=1 +\delta W\ ,\quad H=H_0\left( 1 + \delta h\right)\ .
\ee
Here we assume $|\delta X|$, $|\delta W|$, $|\delta h|\ll 1$. Then from (\ref{A3}), (\ref{A4}), 
and (\ref{A5}), we obtain
\bea
\label{A7}
\frac{d \delta X}{dN} &=& \frac{1}{2}\frac{d \delta h}{dN} - 3 \delta X\ ,\\
\label{A8}
\frac{d \delta W}{dN} &=& - \frac{1}{2}\frac{d \delta h}{dN} - \delta W + 2 \delta X\ ,\\
\label{A9}
\frac{d \delta h}{dN} &=& 4 \delta X\ .
\eea
By deleting $d \delta h/dN$ by using (\ref{A9}) from (\ref{A7}) and (\ref{A8}), we obtain
\be
\label{A10}
\frac{d \delta X}{dN} = - \delta X\ ,\quad \frac{d \delta W}{dN} = - \delta W\ ,
\ee
which tells $\delta X$, $\delta W$ and also $\delta h \propto \e^{-N} \propto 1/a$. 
Therefore the perturbation decreases as the universe expands, which tells that the deSitter solution 
could be stable.

\end{document}